\documentclass[twocolumn]{jpsj2}
\usepackage{color}

\def\runauthor{H. Kobayashi, Y. Fukumoto and A. Oguchi }

\title{%
Frustrated Ising Model on a Diamond Hierarchical Lattice
}

\author{%
Hiroyuki {\sc Kobayashi}, Yoshiyuki {\sc Fukumoto}\thanks{E-mail address: yfuku@ph.noda.tus.ac.jp} and Akihide {\sc Oguchi}
}

\inst{%
Department of Physics, Faculty of Science and Technology, Tokyo University of Science, Noda, Chiba 278-8510
}

\recdate{\hspace{3cm}}

\abst{%
A frustrated Ising model on a diamond hierarchical lattice is studied. 
We obtain the exact partition function of this model and calculate the transition temperature, specific heat, entropy, magnetization, and ferromagnetic correlation function.
Depending on the magnitude of a parameter giving the frustration, there exist three types of ground states: ferromagnetic, classical spin-liquid
with highly developed short-range order, and paramagnetic.
The dependence of the zero-temperature entropy on the frustration parameter has an infinite number of steps.
The temperature dependence of the specific heat exhibits many peaks with decreasing temperature and entropy loss.
The dominant spin configurations at low temperatures are also specified.
}

\kword{%
 classical spin-liquid ground state, frustration, diamond hierarchical lattice, Ising model, phase transition
}

\begin{document}
\maketitle

\section{Introduction}

Many magnetic systems with competing interactions exhibit frustration that leads to multiple ground states called spin glass or spin liquid even at the classical level.  
These interesting properties are attributable to the delicate balance of frustrated spin-spin interactions.
If we can obtain the exact partition function of a model with frustrated magnetic phenomena, we can obtain a better understanding of complicated frustrated phenomena.

In statistical physics, few exactly soluble models with phase transitions are known, such as two-dimensional Ising model~\cite{rf:0}, eight-vertex model~\cite{rf:01},
one-dimensional van der Waals gas model,~\cite{rf:02} and a hierarchical model~\cite{rf:1}. 
Since Berker and Ostlund proposed a hierarchical model related to the renormalization group method~\cite{rf:1,rf:2},
many different models on hierarchical lattices have been proposed and developed.~\cite{rf:2,rf:3,rf:4,rf:5}
The magnetic and thermodynamic behaviors have been studied,~\cite{rf:6,rf:7,rf:8}
and the distribution of the zeros of the partition function and the critical exponents have also been obtained.~\cite{rf:7,rf:9} 
Using a hierarchical model with competing ferro- and antiferromagnetic interactions, McKay {\it et al.} studied the spin-glass behavior~\cite{rf:10}
and Nogueria {\it et al.} investigated the local magnetization.~\cite{rf: 11}
However, to the best of our knowledge, no hierarchical model describes the spin-liquid ground state.
In this study, we consider a hierarchical Ising model with frustrated interactions that lead to the classical spin-liquid ground state. 

This paper is organized as follows.
We introduce our diamond hierarchical lattice and frustrated Ising model in \S 2 and describe our recursion relations in \S 3.
Thermodynamic and ground state properties are described in \S 4 and \S 5, respectively.
In \S 6, the temperature evolution of dominant spin configurations at low temperatures is discussed.
In \S 7, we summarize the results obtained in this study.

\section{Lattice and Hamiltonian}

\begin{figure}[t]
\begin{center} 
\includegraphics[width=.95\linewidth]{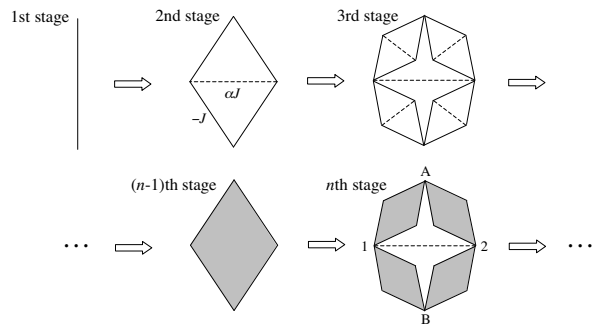}
\end{center}
\caption{Construction of the hierarchical lattice.
The $n$-stage lattice comprises four $(n-1)$-stage lattices
represented by the gray diamonds. The sites denoted by
A, B, 1, and 2 are contact sites between two of the 
$(n-1)$-stage lattices.}
\label{fig:f1}
\end{figure}
  
Our diamond hierarchical lattice is constructed by the infinite iteration procedure shown in Fig.~\ref{fig:f1}.
We begin with a single bond expressed by the solid line, and call this the first stage. 
We replace this single bond by a diamond-shaped basic unit comprising four solid lines and one dotted line to obtain the second-stage lattice.
In order to obtain the third-stage lattice, 
each of the solid lines in the second-stage lattice is replaced by a diamond unit, and the dotted line is left untouched. 
In general, the $n$-stage lattice is constructed by replacing each of the solid lines in the $(n-1)$-stage lattice a the diamond unit.
The dotted lines are always left untouched when we proceed to the next stage.
It should be noted that the $n$-stage lattice comprises four $(n-1)$-stage lattices.
The $n$-stage lattice has four contact sites between the $(n-1)$-stage lattices, and these are indicated by A, B, 1, and 2 in Fig.~\ref{fig:f1}.
We call sites A and B (1 and 2) as surface (internal) contact sites.
The $n$-stage lattice has $N_n=\frac{2}{3}(4^{n-1}+2)$ sites,  $4^{n-1}$ solid lines, and $\frac{1}{3}(4^{n-1}-1)$ dotted lines.
   
In order to define a frustrated Ising model on the diamond hierarchical lattice, we place Ising spins on each of the lattice sites. 
We call spins $\sigma_A$ and $\sigma_B$ defined on the surface contact sites as ``surface spins,"
and spins $\sigma_1$ and $\sigma_2$ defined on the internal contact sites as ``internal contact spins."
Spins on both ends of a solid and a dotted line couple ferromagnetically and antiferromagnetically by $-J$ and $\alpha J$, respectively.
The Hamiltonian can be written as
\begin{equation} 
   {\cal H}=-J\sum_{\langle i,j \rangle} \sigma_i \sigma_j +\alpha J \sum_{\langle\!\langle i,j \rangle\!\rangle}\sigma_i \sigma_j-H\sum_i \sigma_i ,
\end{equation}
where the first and second sums run over all pairs on the solid and dotted bonds, respectively. 
Although we study the zero-field properties, we have introduced the Zeeman term in our Hamiltonian to calculate the magnetization and correlation functions.

Antiferromagnetic coupling tends to destabilize the ferromagnetism induced by the ferromagnetic coupling.
At finite temperatures, there exists a continuous phase transition between the ferromagnetic and disordered phases as a function of $\alpha$.
At zero temperature, the phase transition cannot be continuous because no fluctuations exist.
In this study, we show that an infinite number of first-order phase transitions occur at zero temperature instead of a continuous phase transition at finite temperatures.
Disordered ground states near the ferromagnetic transition point are regarded as spin-liquid because the ferromagnetic short-range order is highly developed.
The residual entropy is step-like as a function of $\alpha$, leading to unique thermodynamic properties.

\section{Recursion Relations}

Here, we derive recursion relations for our hierarchical model.
The partition function of the $n$-stage lattice is given as
\begin{equation}
     Z_n=\sum_{\sigma_A,\sigma_B}Q_n(\sigma_A,\sigma_B)e^{h(\sigma_A+\sigma_B)}, 
\end{equation}
where 
\begin{equation}
     Q_n(\sigma_A,\sigma_B)=A_ne^{K_n\sigma_A\sigma_B+h_n(\sigma_A+\sigma_B)}
\label{eq:3}
\end{equation} 
and $h=H/T$. (We use the unit with $k_{\rm B}=1$ in this paper.)
For $n=1$, we have $A_1=1$, $K_1=K \;(\equiv J/T)$, and $h_1=0$.
For $n \geq 2$, noting that the $n$-stage lattice comprises four $(n-1)$-stage lattices, 
we can express $Q_n(\sigma_A,\sigma_B)$ in terms of the $(n-1)$-stage lattice as follows:
\begin{eqnarray}
   Q_n(\sigma_A,\sigma_B)&&\hspace{-7mm}=\sum_{\sigma_1,\sigma_2} Q_{n-1}(\sigma_A,\sigma_1)Q_{n-1}(\sigma_1,\sigma_B)
\nonumber\\&&\hspace{2mm}
   \times Q_{n-1}(\sigma_A,\sigma_2)Q_{n-1}(\sigma_2,\sigma_B)
\nonumber\\&&\hspace{2mm}
   \times e^{-\alpha K\sigma_1\sigma_2+h(\sigma_1+\sigma_2)}
\nonumber\\  &&\hspace{-20mm}
   =2A_{n-1}^4\{e^{\alpha K}+e^{-\alpha K}
\nonumber\\  &&\hspace{-10mm}
   \times\cosh [2K_{n-1}(\sigma_A+\sigma_B)+4h_{n-1}+2h]\}.\;\;\;
\label{eq:4}
\end{eqnarray}
Using eqs.~(\ref{eq:3}) and (\ref{eq:4}), we obtain the following recursion relations:
\begin{eqnarray}
   K_n && \hspace{-7.5mm}=f_K(K_{n-1},h_{n-1},h)+f_K(K_{n-1},-h_{n-1},-h)
\nonumber   \\&& \hspace{-4.5mm}
   -2f_K(0,h_{n-1},h),
\end{eqnarray}
\begin{eqnarray}
   h_n && \hspace{-7.5mm}=2h_{n-1}+f_K(K_{n-1},h_{n-1},h)
\nonumber   \\&& \hspace{-4.5mm}
   -f_K(K_{n-1},-h_{n-1},-h),
\end{eqnarray}
and 
\begin{equation}
   \ln A_n =4\ln A_{n-1}+K_n+2f_K(0,h_{n-1},h),
\end{equation}
where 
\begin{equation}
   f_K(\bar{K},\bar{h},h)=\frac{1}{4}\ln [e^{\alpha K}+e^{-\alpha K} \cosh(4\bar{K}+4\bar{h}+2h)].
\end{equation}

We express $K_i$, $h_i$, and $\ln A_i \equiv \tilde{A}_i$ as a power series in $h$ as follows:
\begin{equation}
   K_i=K_i^{(0)}+K_i^{(2)}h^2+{\cal{O}}(h^4),
\end{equation}  
\begin{equation}
   h_i=h_i^{(1)}h+{\cal{O}}(h^3),
\end{equation}  
\begin{equation}
   \tilde{A}_i=\tilde{A}_i^{(0)}+\tilde{A}_i^{(2)}h^2+{\cal{O}}(h^4).
\end{equation}  
The recursion relation for $K_n^{(0)}$ is given by
\begin{equation}
   K_n^{(0)} = \frac{1}{2}\ln \frac{e^{\alpha K}+e^{-\alpha K} \cosh(4K_{n-1}^{(0)})}{2\cosh(\alpha K)}.
\label{eq:13}
\end{equation}
We can write $\tilde{A}_n^{(0)}$ and $h_n^{(1)}$ in terms of $K_n^{(0)}$ as follows:
\begin{equation}
   \tilde{A}_n^{(0)}=\sum_{i=2}^{n} 4^{n-i}K_i^{(0)}+\frac{4^{n-1}-1}{3} \ln [4\cosh(\alpha K)],
\label{eq:14}
\end{equation}
\begin{equation}
   h_n^{(1)}=\sum_{i=1}^{n-2}p_i \prod_{s=i+1}^{n-1}2(1+p_s)+p_{n-1},
\label{eq:15}
\end{equation}
where
\begin{equation}
   p_i = \frac{e^{-\alpha K} \sinh(4K_{i}^{(0)})}{e^{\alpha K}+e^{-\alpha K} \cosh(4K_{i}^{(0)})}.
\end{equation}
In a similar manner, we have
\begin{equation}
   K_n^{(2)}=\sum_{i=1}^{n-2}q_i \prod_{s=i+1}^{n-1}2p_s+q_{n-1}
\label{eq:17}
\end{equation}
with
\begin{eqnarray}
   q_i && \hspace{-7mm}=\left\{\frac{\tanh(\alpha K)-1}{2}
   +\frac{e^{2\alpha K}\cosh(4K_{i}^{(0)})+1}{[\cosh(4K_{i}^{(0)})+e^{2\alpha K}]^2}\right\}
\nonumber   \\&& \hspace{0mm}
   \times (1+2h_i^{(1)})^2,
\end{eqnarray}
and
\begin{equation}
   \tilde{A}_n^{(2)}=\sum_{i=2}^{n}4^{n-i}\{K_{i}^{(2)}-(1+h_{i-1}^{(1)})^2[\tanh(\alpha K)-1]\}.
\label{eq:19}
\end{equation}
We use $K_n^{(0)}$, $K_n^{(2)}$, $h_n^{(1)}$, $\tilde{A}_n^{(0)}$, and $\tilde{A}_n^{(2)}$ obtained from 
eqs.~(\ref{eq:13})--(\ref{eq:19}) to study the phase diagram, specific heat, entropy, magnetization, and ferromagnetic correlation function in the next section.
When calculating these quantities,
we choose the maximum value of $n$ as $20\!\sim\!25$ and check if the calculated results can be regarded as those in the thermodynamic limit.

\section{Thermodynamic Properties}

\subsection{Phase diagram}

\begin{figure}[h]
\begin{center}
\includegraphics[width=.8\linewidth]{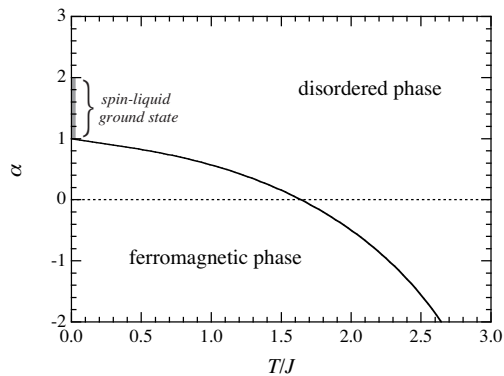}
\end{center}
\caption{Phase diagram in the $\alpha-T$ plane.
At zero temperature, a short-range ferromagnetic order develops
for $1<\alpha<2$ (spin-liquid ground state) but not for 
$2<\alpha$ (paramagnetic ground state).}
\label{fig:f2}
\end{figure}

The transition temperature and phase diagram can be determined from the fixed point of eq.~(\ref{eq:13}).
The phase boundary in the $\alpha-T$ plane is given by
\begin{equation}
   \alpha = \frac{T}{2J} \ln \left(\frac{e^{\frac{2J}{T}}-e^{-\frac{2J}{T}}-e^{-\frac{4J}{T}}-1}{2}\right).
\end{equation}
The phase diagram is shown in Fig.~\ref{fig:f2}.
When $\alpha <1$, a second-order transition  from the ferromagnetic phase to the disordered phase occurs at finite  temperature.
For $1<\alpha<2$, it will be shown later that the ground state becomes a classical spin-liquid state in which the short-range order is highly developed 
but the magnetization is zero and the entropy is finite.
When $\alpha>2$, the ground state is paramagnetic in which the short-range order vanishes.
The paramagnetic ground state consists of independent dimers composed by the coupling $\alpha J$. 

\subsection{Specific heat and entropy}

Using the relation 
\begin{equation}
    \lim_{h\rightarrow 0}Z_n=4 A_n^{(0)} \cosh K_n^{(0)},
\end{equation}
we obtain the specific heat per spin  
\begin{equation}
   C=\lim_{n\rightarrow \infty}\frac{K^2}{N_n}\frac{\partial^2}{\partial K^2}[\tilde{A}_n^{(0)}+\ln(\cosh K_n^{(0)})],
\end{equation}  
and the entropy per spin     
\begin{equation}
   S=\lim_{n\rightarrow \infty}\frac{1}{N_n}\left(1-K\frac{\partial}{\partial K}\right)[\tilde{A}_n^{(0)}+\ln(\cosh K_n^{(0)})].
\end{equation} 
Using the recursion relations for $K_n^{(0)}$ and $A_n^{(0)}$ in eqs.~(\ref{eq:13}) and (\ref{eq:14}),
we can evaluate $C$ and $S$ at finite temperatures.

\begin{figure}[t]
\begin{center}
\includegraphics[width=.8\linewidth]{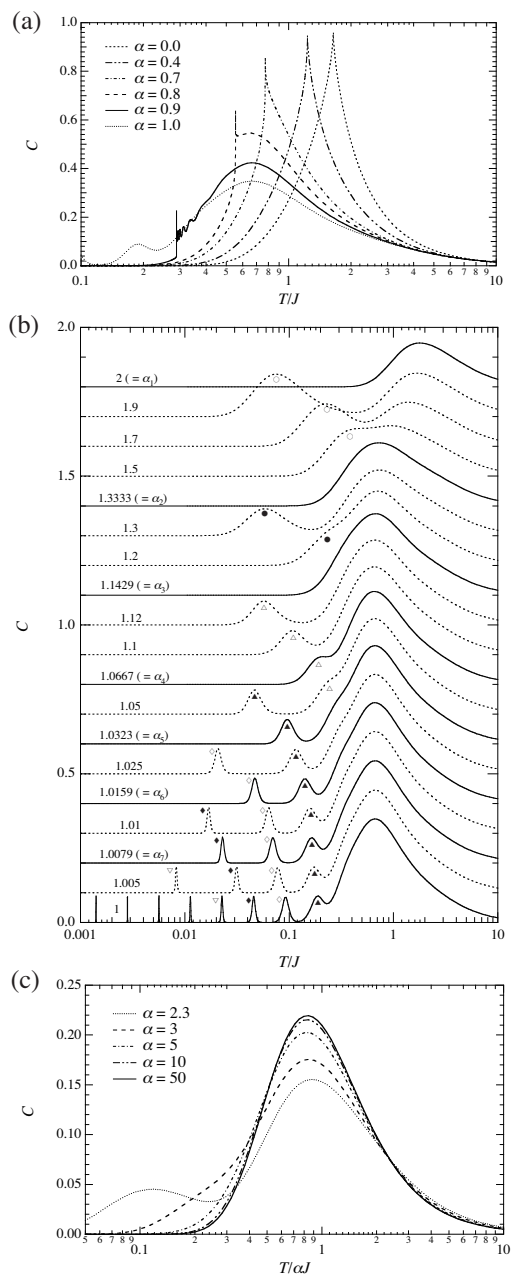}
\end{center}
\caption{Temperature dependence of the specific heat
for (a) $0\leq \alpha \leq 1$, (b) $1 \leq \alpha \leq 2$, and (c) $\alpha > 2$.
In (b), the original curves are shifted upwards by multiples of $0.1$, and
the number corresponding to each curve represents the value of $\alpha$.}
\label{fig:f3}
\end{figure}

\begin{figure}[t]
\begin{center}
\includegraphics[width=.8\linewidth]{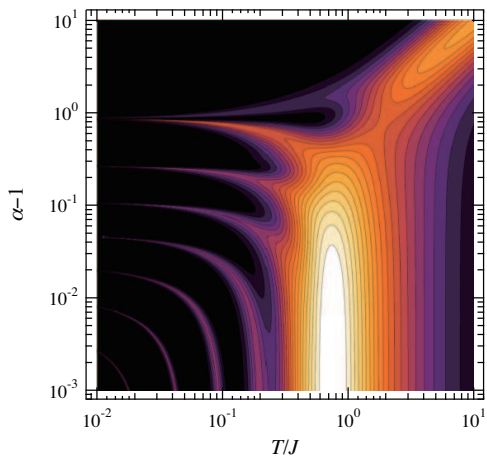}
\end{center}
\caption{Contour map of the specific heat in the $\alpha-T$ plane
for $\alpha>1$.}
\label{fig:C_cntr}
\end{figure}

The dependence of the specific heat on temperature is shown in Fig.~\ref{fig:f3}.
The results for $\alpha=0\!\sim\! 1$, for which a ferromagnetic phase appears at low temperatures, are shown in Fig.~\ref{fig:f3}(a).
The specific heat for $\alpha=0$ diverges at the ferromagnetic transition temperature $T_{\rm c}=1.641 J$.~\cite{rf:6}
When $\alpha$ increases, the divergence point decreases, which agrees with the behavior of the above-obtained phase boundary.
For $\alpha=0.8$, we find a broad maximum at $T=0.66 J$, and the ferromagnetic transition temperature $T_{\rm c}=0.5545 J$ is lower than the broad maximum temperature.
For $\alpha=0.9$, many peaks appear at temperatures slightly greater than the transition temperature $T_{\rm c}=0.2881J$.
Figure~\ref{fig:f3}(b) shows the specific heat for $\alpha=1 \sim 2$.
We find that the result for $\alpha=1$ has an infinite number of broad peaks.
For $\alpha>1$, the number of broad peaks becomes finite. 
The evolution of the specific heat curve as a function of $\alpha$ exhibits a complex behavior, 
and it is described in detail in relation to the entropy later.
Figure~\ref{fig:f3}(c) shows the results for $\alpha>2$.
In this figure, we choose $\alpha J$ as the unit of temperature.
The result for $\alpha=2.3$ has two peaks.
When $\alpha$ is increased, the low-temperature peak merges with the high-temperature one.
Then, with a further increase in $\alpha$, the curve moves toward the $\alpha=\infty$ limit monotonically.
In Fig.~\ref{fig:C_cntr}, we show the results shown in Figs.~\ref{fig:f3} (b) and (c) in the form of a contour map in the $\alpha-T$ plane.

\begin{figure}[b]
\begin{center}
\includegraphics[width=.8\linewidth]{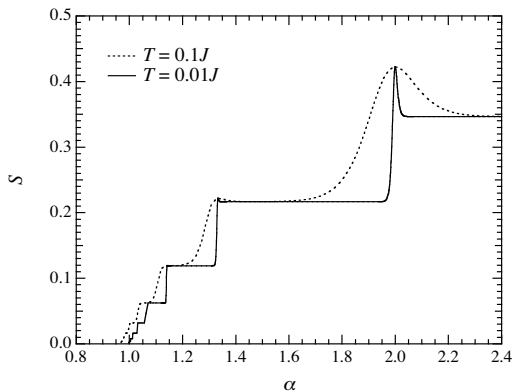}
\end{center}
\caption{Entropy $S$ as a function of $\alpha$ at $T=0.1J$ and $0.01J$}
\label{fig:f4}
\end{figure}

It is helpful to calculate the entropy in order to understand the above mentioned behavior of the specific heat.
The entropy per spin as a function of $\alpha$ at low temperatures is shown in Fig.~\ref{fig:f4}.
We have $S=0$ in the ferromagnetic ground state for small $\alpha$ and $S=(1/2)\log 2\simeq 0.347$ in the paramegnetic ground state for large $\alpha$ at $T \rightarrow 0$.
We find that many steps appear between them.
We define the positions of steps at $T \rightarrow 0$ as $\alpha_1$, $\alpha_2$, $\cdots$ in descending order.
(In the next section, it is shown that $\alpha_l=2^l/(2^l-1)$.)
When $\alpha$ decreases and passes $\alpha_l$, an additional peak appears in the temperature dependence of the specific heat, as seen in Fig.~\ref{fig:f3}(b).
Figure \ref{fig:C_cntr} clearly indicates that we can ascribe each low-temperature peak in the specific heat to a step in the $\alpha$-dependence of entropy.
For example, in Fig.~\ref{fig:f3}(b), the peaks (or shoulders) of $\alpha=1.9$, 1.7, 1.5 indicated by the open circles correspond to
the entropy step at $\alpha=\alpha_1$ $(=2)$,
and those of $\alpha=1.3$, 1.2 indicated by the closed circles correspond to the entropy step at $\alpha=\alpha_2$ $(=4/3)$.
The entropy exhibits a spike at around $\alpha=2$, as seen in Fig.~\ref{fig:f4}.
Then, the specific heat for a slightly greater value of $\alpha$ may have an additional peak.
In fact, the specific heat curve for $\alpha=2.3$ shown in Fig.~\ref{fig:f3}(c) has an additional low-temperature peak.
The next and more important problem is why the $\alpha$-dependence of entropy has such a structure. 
This issue is discussed in the following section.

\subsection{Magnetization}

Griffiths and Kaufman proved that the hierarchical  model has a thermodynamic limit.~\cite{rf:3}  
The surface spin state does not have a significant effect on the physical properties of the system in the limit of $n \rightarrow \infty$. 
Therefore, instead of an extremely small external field applied on this system, we use $\sigma _A=1$ and $\sigma _B=1$.
The magnetization per site, $m$, is defined by following expression:
\begin{eqnarray}
   m && \hspace{-7mm} = \lim_{n\rightarrow\infty}\frac{1}{N_n}
   \lim_{h\rightarrow 0}\frac{\partial}{\partial h}\ln Q_n(\sigma_A=1,\sigma_B=1)
\nonumber   \\&& \hspace{-7mm}
   =\lim_{n\rightarrow\infty}\frac{3h_n^{(1)}}{4^{n-1}+2}.
\end{eqnarray}
If the system does not have a long-range order, no magnetization appears under this setting of boundary surface,
and if it does, a finite magnetization can be obtained in the limit of a vanishing external magnetic field.

\begin{figure}[h]
\begin{center}
  \includegraphics[width=.8\linewidth]{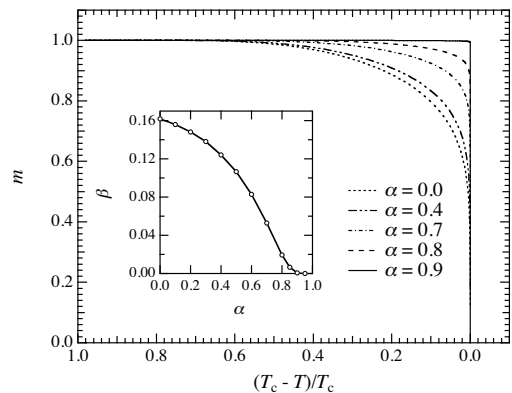}
\end{center}
\caption{Magnetization as a functions of temperature.
The inset shows the $\alpha$-dependence of the critical
index $\beta$, defined by $m \propto (T_{\rm c}-T)^{\beta}$
around $T=T_{\rm c}$.}
\label{fig:f5}
\end{figure}

The temperature dependence of magnetization for $0 \leq \alpha <1$ is shown in Fig.~\ref{fig:f5},
where the horizontal axis represents the reduced temperature, $(T_{\rm c}-T)/T_{\rm c}$.
For $\alpha=0$, we obtain a magnetization index $\beta=0.1617$, which agrees with the result obtained by Margado {\it et al.}~\cite{rf:6}
With a decrease in $\alpha$, the form of the magnetization curve tends to resemble a step function, and the index $\beta$ decreases and vanishes for $\alpha\rightarrow 1$
(see the inset in Fig.~\ref{fig:f5}).
The continuous variation of the critical exponents has been reported in previous studies.~\cite{rf:3,rf:4}

\subsection{Ferromagnetic correlation function}

We calculate a ferromagnetic correlation function to study how the short-range order develops in the disordered region. 
Using  
\begin{equation}
    \Delta_{\rm F}(n)=\frac{1}{\sqrt{N_n}}\sum_i \sigma_i,
\end{equation}
we define the ferromagnetic correlation function as
\begin{eqnarray}
    \langle \Delta_{\rm F}^2 \rangle && \hspace{-7mm} \equiv \lim_{n \rightarrow \infty} \langle \Delta_{\rm F}^2(n) \rangle
    \nonumber \\&& \hspace{-7mm}
    =\lim_{n \rightarrow \infty}\frac{1}{N_n}\lim_{h\rightarrow 0} \frac{\partial^2 \ln Z_n}{\partial h^2}
    \nonumber \\&& \hspace{-7mm}
    =\left.\lim_{n \rightarrow \infty}\frac{3}{4^{n-1}+2}\right[\tilde{A}_n^{(2)}+K_n^{(2)}
    \nonumber \\&& \hspace{8mm}\left.
    +2\frac{(1+h_n^{(1)})^2-e^{-2K_n^{(0)}}K_n^{(2)}}{1+e^{-2K_n^{(0)}}}
    \right].
\label{eq:26}
\end{eqnarray}

\begin{figure}[h]
\begin{center}
  \includegraphics[width=.8\linewidth]{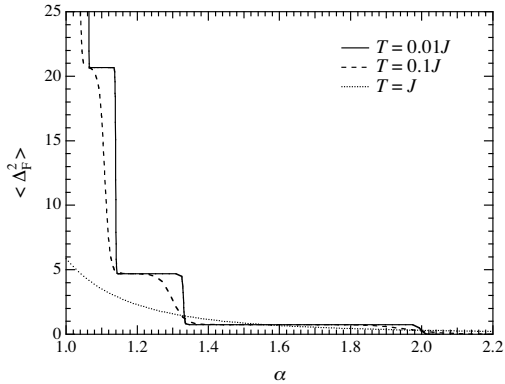}
  \end{center}
\caption{Ferromagnetic correlation function as a function of 
$\alpha$ at $T=J$, $0.1J$, and $0.01J$.}
\label{fig:f6}
\end{figure}

In Fig.~\ref{fig:f6}, we show the $\alpha$-dependence of the correlation function at $T=0.01 J$, $0.1J$, and $J$.
A system with $\alpha>1$ is in the disordered phase at all temperatures.
However, if $\alpha \sim 1$, the short-range ferromagnetic correlation is well developed at low temperatures.
We have $\lim_{T \rightarrow 0}\langle\Delta_{\rm F}^2\rangle>0$ for $1<\alpha<2$ and $\lim_{T \rightarrow 0}\langle\Delta_{\rm F}^2\rangle=0$ for $\alpha>2$.

A distinctive feature of the present model is that a short-range order develops for $1<\alpha<2$ between the ferromagnetic and paramagnetic ground states. 
If the ferromagnetic state for $\alpha<1$ is comparable to the solid state of matter and the paramagnetic state for $\alpha>2$ to the gas state, 
this state for $1<\alpha<2$ is comparable to the liquid state. 
Therefore, we can consider this ground state in the region of $1<\alpha <2$ as a classical spin-liquid ground state.

\section{Ground State Properties}

In the previous section, we have shown that a spin-liquid ground state appears between the ferromagnetic
and paramagnetic ground states on the basis of the calculated results of the entropy and the correlation function.
Here, we study ground-state spin configurations in order to understand how the short-range order develops in the spin-liquid ground states.

\subsection{Ground state energy and spin configurations}

We begin with the case $n=2$, and denote the spin configuration as $(\sigma_A, \sigma_B,\sigma_1,\sigma_2)$.
When $\alpha$ is small, the lowest-energy spin configurations are ferromagnetic states, 
$(\uparrow, \uparrow, \uparrow, \uparrow)$ and $(\downarrow, \downarrow, \downarrow, \downarrow)$,
and the ground state energy is given by $-4+\alpha$.
When $\alpha$ increases, the ferromagnetic states are destabilized against eightfold-degenerate spin configurations,
$\{(\sigma_A, \sigma_B,\sigma_1,\sigma_2)|\sigma_A=\uparrow,\downarrow;\sigma_B=\uparrow,\downarrow;\sigma_1=\uparrow,\downarrow;\sigma_2=\sigma_1\}$,
whose energy is $-\alpha$. The energy crossing point is $\alpha=2\; (\equiv\alpha_1)$.
Therefore, the ground state energy of the second-stage lattice, $E_2(\alpha)$, is given by
\begin{equation}
    E_2(\alpha)=\left\{
    \begin{array}{ll}
       -\alpha & \mbox{for $\alpha_1<\alpha$}\\
       -4+\alpha & \mbox{for otherwise}
    \end{array}
    \right..
\end{equation}
Thus, the ground state for $n=2$ is divided into two sectors depending on $\alpha$.
We call $\alpha_1<\alpha$ as the first sector and the other one as the ferromagnetic sector.
In addition, the ground state energy of the first (ferromagnetic) sector of the second-stage lattice is denoted by $E_{2,1}\equiv -\alpha$ ($E_{2,f}\equiv -4+\alpha$).

\begin{figure}[b]
\begin{center}
\includegraphics[width=.75\linewidth]{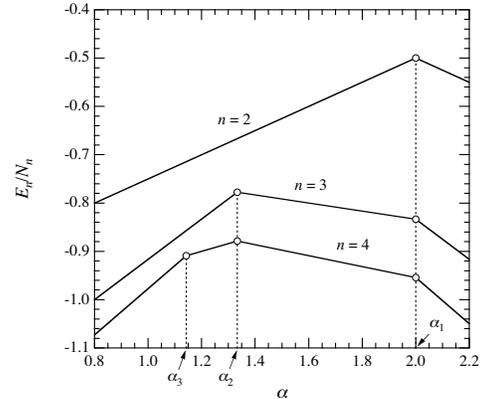}
\end{center}
\caption{Dependence of the ground state energy per site on
$\alpha$ for $n=2$, 3, 4.}
\label{fig:f7}
\end{figure}

For the third-stage lattice, there are two crossing points in the ground state energy $E_3(\alpha)$ as a function of $\alpha$ (see Fig.~\ref{fig:f7}),
and thus, we have three sectors: the first sector is $\alpha_1<\alpha$; the second, $\alpha_2<\alpha<\alpha_1$, where $\alpha_2=4/3$;
and the third, $\alpha<\alpha_2$, is the ferromagnetic sector .
The energy of the first sector, $E_{3,1}=-5\alpha$, satisfies $E_{3,1}=4E_{2,1}-\alpha$,
which shows that each of the four diamonds in the third-stage lattice gives energy $E_{2,1}$ and the internal contact spins $\sigma_1$ and $\sigma_2$ couple antiferromagnetically.
For the energy of the ferromagnetic sector, $E_{3,f}=-16+5\alpha$, we have $E_{3,f}=4E_{2,f}+\alpha$.
Note that the internal contact spins in the ferromagnetic sector lead to energy loss by $\alpha$.
This energy loss is the cause of the appearance of the second sector in the third-stage lattice.
For the energy of the second sector, $E_{3,2}=-8-\alpha$, the relationship $E_{3,2}=2E_{2,1}+2E_{2,f}-\alpha$ holds.
This relationship indicates that two of the four diamonds give up being in the ferromagnetic state to obtain an energy gain from the antiferromagnetic interaction between $\sigma_1$ and $\sigma_2$.

For the fourth-stage lattice, we have four sectors, as seen in Fig.~\ref{fig:f7}.
It is easy to check that recursion relations for energies similar to the third-stage lattice also hold.
In general, we have $n$ sectors in the ground state of the $n$-stage lattice.
We denote the ground state energy of each sector by $E_{n,1}$ for $\alpha_1<\alpha$, $E_{n,2}$ for $\alpha_2<\alpha<\alpha_1$, $\cdots$,
$E_{n,n-1}$ for $\alpha_{n-1}<\alpha<\alpha_{n-2}$, and $E_{n,f}$ for $\alpha<\alpha_{n-1}$.
It is expected that the $(n-1)$-th sector comprises the $(n-2)$-th and ferromagnetic sectors in the $(n-1)$-stage lattice,
and each of the other sectors comprises sectors from the same number in the $(n-1)$-stage lattice;
this is confirmed in the following subsections by comparing the zero-temperature entropy and correlation function with those obtained by the finite-temperature calculation presented in the previous section.

Here, we derive expressions of the ground state energy for the $n$-stage lattice on the basis of the above observation.
For the ferromagnetic sector, we have
\begin{equation}
    E_{n,f}=4E_{n-1,f}+\alpha.
\end{equation}
This recursion relation, along with the initial condition $E_{2,f}=-4+\alpha$, gives
\begin{equation}
    E_{n,f}=-4^{n-1}+\frac{\alpha}{3}(4^{n-1}-1).
\end{equation}
For sector $l=n-1$, using the recursion relation
\begin{equation}
    E_{n,n-1}=2E_{n-1,f}+2E_{n-1,n-2}-\alpha
\end{equation}
and the initial condition $E_{2,1}=-\alpha$, we obtain
\begin{equation}
    E_{n,n-1}=2^n-4^{n-1}+\frac{\alpha}{3}(4^{n-1}-3\cdot 2^n+5).
\label{eq:40}
\end{equation}
Finally, for sectors $l=1,\cdots, n-2$, the recursion relation
\begin{equation}
    E_{n,l}=4E_{n-1,l}-\alpha
\end{equation}
gives
\begin{eqnarray}
    E_{n,l}&&\hspace{-7mm}=2^{2n-l-1}-4^{n-1}
    \nonumber \\ &&\hspace{-4mm}
    +\frac{\alpha}{3}(4^{n-1}-3\cdot 2^{2n-l-1}+4^{n-l}+1),
\end{eqnarray}
where we have used the expression of $E_{l+1,l}$ obtained from eq.~(\ref{eq:40}).
The condition $E_{n,l+1}=E_{n,l}$ yields the lower bound $\alpha_l$ in the form
\begin{equation}
    \alpha_l=\frac{2^l}{2^l-1}
\end{equation}
for $l=1,2,\cdots,n-1$.

In order to study the lowest-energy spin configurations in greater detail, it is convenient to introduce the following notations.
We represent a spin configuration of the $n$-stage lattice as $(\sigma_A, \sigma_B, s)$, where $\sigma_A$ and $\sigma_B$ are surface spins and $s$ represents the configuration of the other spins.
We define $C_{n,l}^{(\sigma_A,\sigma_B)}$ as a set of internal configurations $s$ that gives the lowest energy for the $l$th sector under the condition
that the surface spin state is fixed as $(\sigma_A,\sigma_B)$.
For example, setting $s=(\sigma_1,\sigma_2)$ for the second stage ($n=2$), we have
\begin{equation}
    C_{2,1}^{(\sigma_A,\sigma_B)}=\{(\uparrow,\downarrow), (\downarrow,\uparrow)\}
\end{equation}
for the first sector ($l=1$) and
\begin{equation}
    C_{2,f}^{(\sigma_A,\sigma_B)}=\left\{
    \begin{array}{cl}
       \{(\sigma_A,\sigma_B)\} & \mbox{for $\sigma_A=\sigma_B$}\\
       \emptyset & \mbox{for $\sigma_A\neq\sigma_B$}
    \end{array}
    \right.
\end{equation}
for the ferromagnetic sector.
A detailed discussion of $C_{n,l}^{(\sigma_A,\sigma_B)}$ can be found in Appendix A.
In particular, the derivation of an analytical expression of $|C_{n,l}^{(\sigma_A,\sigma_B)}|$, 
the total number of elements in the set $C_{n,l}^{(\sigma_A,\sigma_B)}$, is presented there.

\subsection{Entropy}

The zero-temperature entropy per site for the $l$th sector at $n\rightarrow\infty$ can be written as
\begin{eqnarray}
   S_{\infty,l}&&\hspace{-7mm}=\lim_{n\rightarrow\infty}\frac{1}{N_n}\log\left(\sum_{\sigma_A,\sigma_B}|C_{n,l}^{(\sigma_A,\sigma_B)}|\right)
    \nonumber \\ &&\hspace{-7mm}
    =\frac{3\cdot 2^{l-1}-1}{4^l}\log 2,
\end{eqnarray}
where we have used the expression for $|C_{n,l}^{(\sigma_A,\sigma_B)}|$ given in eq.~(\ref{eq:46}) of Appendix A.
Denoting the entropy as a function of $\alpha$ by $S(\alpha)$, we have $S(\alpha)=S_{\infty,l}$ for $\alpha_l<\alpha<\alpha_{l-1}$.
With regard to the entropy at a boundary $\alpha=\alpha_l$, 
noting that $S(\alpha)$ at $T=0.01J$ exhibits at around $\alpha=\alpha_1$ (Fig.~\ref{fig:f4}), we expect that $S(\alpha_l)>S_{\infty,l}$ at $T=0$.
As derived in Appendix B, $S(\alpha_l)$ can be written as
\begin{equation}
    S(\alpha_l)=S_{\infty,l}+\frac{3}{8\cdot 4^l}
    \log\left[1+4(2^{-2^l}+4^{-2^l})\right].
\end{equation}

\begin{figure}[h]
\begin{center}
\includegraphics[width=.75\linewidth]{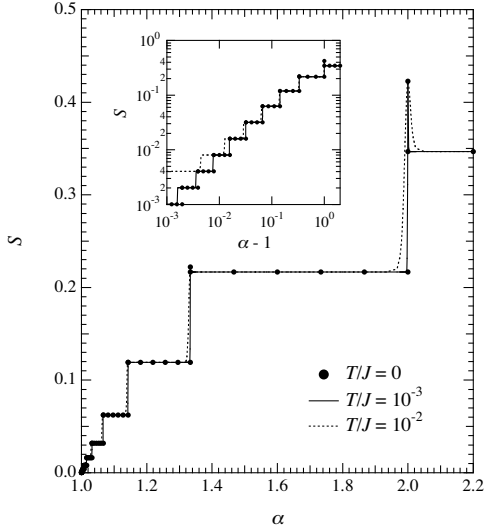}
\end{center}
\caption{Residual entropy as a function of $\alpha$.
The inset show a double logarithmic plot.
}
\label{fig:f10}
\end{figure}

In Fig.~\ref{fig:f10}, we show the entropy obtained by the above expressions, 
along with that obtained by the finite-temperature calculation presented in the previous section.
We find that both results are consistent with each other.
A clear spike is obtained at around $\alpha=\alpha_1 (=2)$.
The height of this spike is $S(\alpha_1)-S_{\infty,1}=\frac{3}{32}\log\frac{9}{4}\simeq 0.076$.
There is a spike at around $\alpha=\alpha_l$ for $l \geq 2$; the height of this spike decreases rapidly with an increase in $l$. 
The inset shows a double logarithmic plot that shows the behavior in the immediate vicinity of $\alpha=1$.
We find that the discrepancy between the results of $T/J=10^{-2}$ and $T/J=0$ is pronounced below $\alpha-1\simeq 10^{-2}$.
However, the result of $T/J=10^{-3}$ indicates that lowering of the temperature resolves the discrepancy, as expected.

\subsection{Magnetization}

In order to obtain the magnetization of the ground state we consider the surface spins to be upward, as before.
We denote the magnetization of spin configurations in $C_{n,l}^{(\uparrow,\uparrow)}$ as $M(C_{n,l}^{(\uparrow,\uparrow)})$.
Then, we find $M(C_{2,1}^{(\uparrow,\uparrow)})=0$ and $M(C_{2,f}^{(\uparrow,\uparrow)})=2$, and obtain the following recursion relations:
\begin{equation}
    M(C_{n,f}^{(\uparrow,\uparrow)})=2+4M(C_{n-1,f}^{(\uparrow,\uparrow)})
\end{equation}
and
\begin{equation}
    M(C_{n,l}^{(\uparrow,\uparrow)})=2M(C_{n-1,l}^{(\uparrow,\uparrow)})
    \hspace{5mm}\mbox{for $l<n$},
\end{equation}
where $C_{n-1,n-1}^{(\uparrow,\uparrow)}\equiv C_{n-1,f}^{(\uparrow,\uparrow)}$.
We have used eqs. (\ref{eq:cnf}), (\ref{eq:33}), and (\ref{eq:cnl}) of Appendix A in the derivation of these relations.
Solving these equations, we have
\begin{equation}
    M(C_{n,l}^{(\uparrow,\uparrow)})=\frac{2^{n-l+1}}{3}(4^{l-1}-1).
\label{eq:59}
\end{equation}
The magnetization per spin for the $l$th sector of the $n$th stage, $m_{n,l}$, is written as
\begin{equation}
    m_{n,l}=\frac{2+M(C_{n,l}^{(\uparrow,\uparrow)})}{N_n}=\frac{3+2^{n-l}(4^{l-1}-1)}{4^{n-1}+2}.
\end{equation}

\begin{figure}[h]
\begin{center}
\includegraphics[width=.82\linewidth]{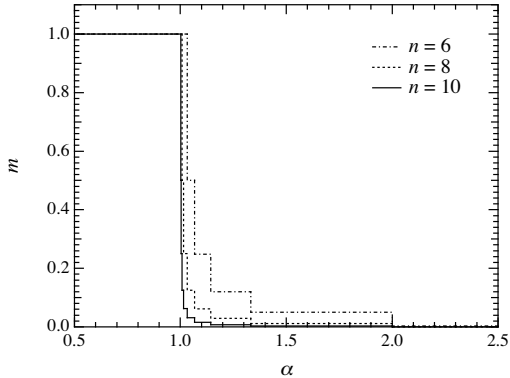}
\end{center}
\caption{Magnetization at $T=0$ as a function of $\alpha$
for finite size systems with $n=6, 8, 10$.}
\label{fig:f11}
\end{figure}

We show the $\alpha$ dependence of magnetizations for $n=6$, 8, 10 in Fig.~\ref{fig:f11}.
We have $m=1$ for $\alpha<1$ and $m=0$ for $\alpha>2$.
For the region $1<\alpha<2$, the magnetization is finite in finite size systems, but it vanishes rapidly with an increase in the system size.
Thus, we have $m=0$ for $1<\alpha<2$ in the thermodynamic limit.

\subsection{Ferromagnetic correlation function}

We consider the ferromagnetic correlation function in the ground state of the $n$-stage lattice.
Denoting the average in the ground-state manifold as $\langle\cdots\rangle_{\rm g}$, we have
\begin{equation}
    \langle\Delta_{\rm F}^2(n)\rangle_{\rm g}=\frac{1}{4N_n}\sum_{\sigma_A,\sigma_B}\left[\sigma_A+\sigma_B+M(C_{n,l}^{(\sigma_A,\sigma_B)})\right]^2
\end{equation}
for $\alpha_l<\alpha<\alpha_{l-1}$.
Using eq.~(\ref{eq:59}), $M(C_{n,l}^{(\downarrow,\downarrow)})=-M(C_{n,l}^{(\uparrow,\uparrow)})$, 
and $M(C_{n,l}^{(\uparrow,\downarrow)})=M(C_{n,l}^{(\downarrow,\uparrow)})=0$, we obtain
\begin{equation}
    \langle\Delta_{\rm F}^2(n)\rangle_{\rm g}=\frac{1}{2N_n}\left[2+\frac{2^{n-l+1}}{3}(4^{l-1}-1)\right]^2.
\end{equation}
In the thermodynamic limit, the correlation function of the $l$th sector is obtained as follows:
\begin{equation}
    \langle\Delta_{\rm F}^2\rangle_{\rm g}=\lim_{n\rightarrow\infty}\langle\Delta_{\rm F}^2(n)\rangle_{\rm g}=\frac{4^l(1-4^{1-l})^2}{12}.
\label{eq:63}
\end{equation}

\begin{figure}[h]
\begin{center}
\includegraphics[width=.78\linewidth]{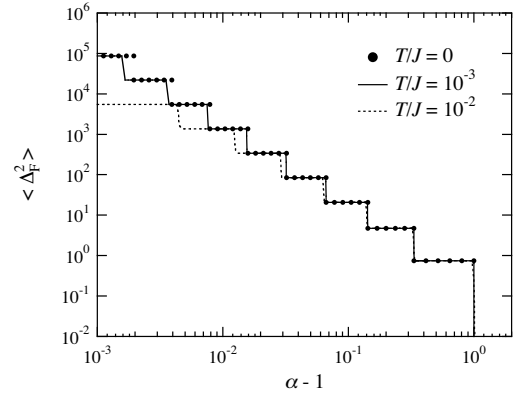}
\end{center}
\caption{Log-log plot of the ferromagnetic correlation function 
at $T=0$ as a function of $\alpha$.}
\label{fig:f12}
\end{figure}

In Fig.~\ref{fig:f12}, we compare the ground state correlation function given by eq.~(\ref{eq:63}) with
the low-temperature limit obtained by eq.~(\ref{eq:26}).
The agreement between both results is reasonable.
The short-range order grows exponentially as a function of sector number $l$.
The most characteristic feature of the present system is the highly developed short-range ordering in the intermediate phases,
although there are no thermal and quantum fluctuations in the classical Ising system at $T=0$.

\section{Temperature Evolution of Dominant Spin Configurations}

Here, we study the temperature evolution of dominant spin configurations at low temperatures for the most characteristic case of $\alpha=1$
by considering the temperature dependence of the entropy and the correlation function.

In the previous section, we have discussed the origin of the step structure of the residual entropy for $1\leq \alpha \leq 2$,
and spin configurations on each of the plateaus have been specified.
We denote the set of the lowest-energy spin configurations of the $l$th sector in the thermodynamic limit, $n\rightarrow\infty$, as $C_l$.
On the one hand, in \S4, we have found that the low-temperature peaks in the specific heat for $\alpha=1$ are related to the steps of the residual entropy as a function of $\alpha$.
Therefore, we expect that dominant spin configurations between two successive low-temperature peaks can be specified in terms of $\{C_l\}$.

\begin{figure}[t]
\begin{center}
\includegraphics[width=.77\linewidth]{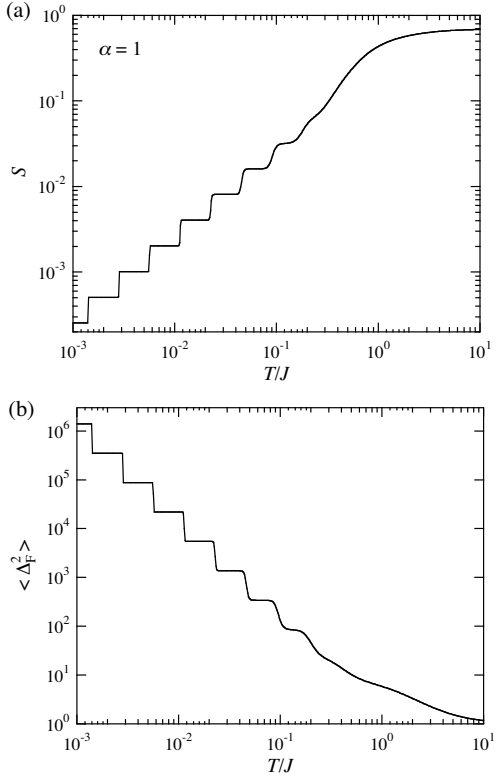}
\end{center}
\caption{Temperature dependence of (a) entropy and (b) correlation 
function for $\alpha=1$.}
\label{fig:f13a}
\end{figure}

\begin{figure}[h]
\begin{center}
\includegraphics[width=.65\linewidth]{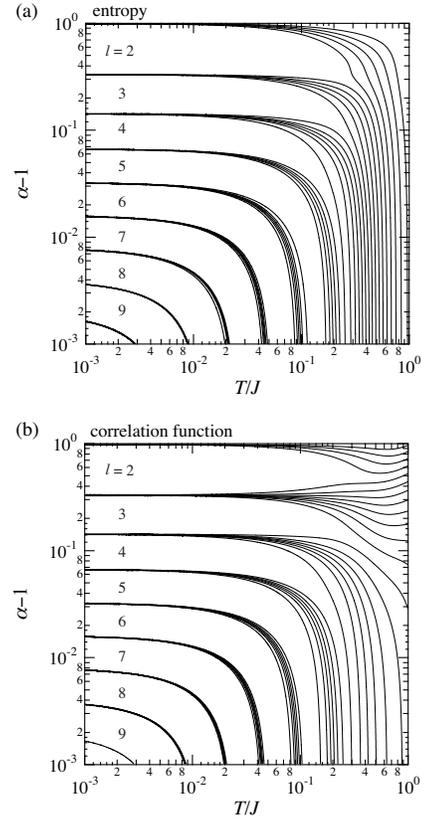}
\end{center}
\caption{Contour plots of (a) $\log S$ and 
(b) $\log \langle\Delta_{\rm F}^2\rangle$ in the $\alpha$-$T$ plane,
where the number corresponding to each region represents the sector number $l$.}
\label{fig:f13b}
\end{figure}

In Fig.~\ref{fig:f13a}, we show the temperature dependence of the entropy $S$ and correlation function $\langle\Delta_{\rm F}^2\rangle$ for $\alpha=1$.
We find that a step-like structure appears in both curves at $T/J<\;\sim\!0.14$.
In order to specify the spin configurations on each of the plateaus, we present contour plots of the entropy and correlation function in Fig.~\ref{fig:f13b}.
As seen from these contour plots, the dominant spin configurations for $0.10<T/J<0.14$ are those in $C_5$.
With a farther decrease in temperature, the dominant spin configurations vary as $C_5 \rightarrow C_6 \rightarrow C_7 \rightarrow C_8 \rightarrow \cdots$.

\section{Conclusions}

We have proposed a frustrated Ising hierarchical model with a diamond-shaped basic unit comprising
four ferromagnetic bonds and one non-iterated diagonal bond,
and we have obtained a rigorous partition function.
From this partition function, we have found that this model has three types of ground states depending on the magnitude of the  frustration parameter $\alpha$. 
The ground state is ferromagnetic for $\alpha < 1$ and paramagnetic with a vanishing ferromagnetic correlation function for $\alpha > 2$.
An intermediate phase appears for $1<\alpha<2$ in our ground state phase diagram, 
and the ground-state and thermal properties exhibit many interesting phenomena in this region.
When $\alpha$ decreases from $2$ to $1$, 
the system goes through an infinite number of first-order transitions to develop a ferromagnetic short-range order:
the ferromagnetic correlation function of the ground-state manifold exhibits step-like increments and diverges, 
and the zero-temperature entropy exhibits step-like decrements and vanishes.
Because the short-range order is highly developed in each set of the lowest-energy spin configurations for $1<\alpha<2$, 
we have identified the intermediate phase to be classical spin-liquid.
For $\alpha \rightarrow 1$, 
there are temperature regions in which one of these sets of spin configurations becomes the dominant one.

In conclusion, the frustrated Ising model on the diamond hierarchical lattice provides a new prototypical example to describe frustrated phenomena.

\appendix
\section{Lowest-Energy Spin Configurations and Degeneration Factor}

Here, we study $C_{n,l}^{(\sigma_A,\sigma_B)}$ for a general value of $n\;(\geq3)$.
It is convenient to write an internal configuration as $s=(s_{A1},s_{B1},\sigma_1,s_{A2},s_{B2},\sigma_2)$, 
where $s_{\eta j}$ denotes a spin configuration of internal spins between the contact sites $\eta$ ($=A$ or $B$) and $j$ ($=1$ or $2$).

For the ferromagnetic sector, setting $\sigma_A=\sigma_B\equiv\sigma$, we simply obtain
\begin{eqnarray}
    C_{n,f}^{(\sigma,\sigma)}&&\hspace{-6mm}=\left\{s|s_{A1},s_{B1}\in C_{n-1,f}^{(\sigma,\sigma)};\sigma_1=\sigma;\right.
    \nonumber \\ &&\hspace{3mm}
    \left.s_{A2},s_{B2}\in C_{n-1,f}^{(\sigma,\sigma)};\sigma_2=\sigma\right\}.
\label{eq:cnf}
\end{eqnarray}

\begin{figure}[h]
\begin{center}
\includegraphics[width=1\linewidth]{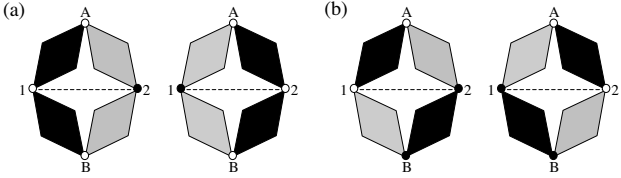}
\end{center}
\caption{Schematic representation of spin configurations in (a) $C_{n,n-1}^{(\sigma,\sigma)}$ and 
(b) $C_{n,n-1}^{(\sigma,-\sigma)}$. The open (closed) circles indicate that these spins take the
value of $\sigma$ ($-\sigma$). The black diamond represents an element in
$C_{n-1,f}^{(\sigma,\sigma)}$ or $C_{n-1,f}^{(-\sigma,-\sigma)}$, and 
the gray diamond represents one that in $C_{n-1,n-2}^{(\sigma,-\sigma)}$ or 
$C_{n-1,n-2}^{(-\sigma,\sigma)}$.}
\label{fig:f8}
\end{figure}

Next, we consider the sector $l=n-1$.
In this case, $C_{n,n-1}^{(\sigma_A,\sigma_B)}$ are constructed by the $(n-1)$-stage ground states of the ferromagnetic and $(n-2)$-th sectors.
For $\sigma_A=\sigma_B \equiv \sigma$, we obtain
\begin{eqnarray}
    C_{n,n-1}^{(\sigma,\sigma)}
    &&\hspace{-7mm}=\left\{s|s_{A1},s_{B1}\in C_{n-1,f}^{(\sigma,\sigma)};\sigma_1=\sigma;\right.
    \nonumber \\ &&\hspace{2mm}
    \left.s_{A2},s_{B2}\in C_{n-1,n-2}^{(\sigma,-\sigma)};\sigma_2=-\sigma\right\}
    \nonumber \\ &&\hspace{-7mm}
    +\left\{s|s_{A1},s_{B1}\in C_{n-1,n-2}^{(\sigma,-\sigma)};\sigma_1=-\sigma;\right.
    \nonumber \\ &&\hspace{2mm}
    \left.s_{A2},s_{B2}\in C_{n-1,f}^{(\sigma,\sigma)};\sigma_2=\sigma\right\}.
\label{eq:33}
\end{eqnarray}
For $\sigma_A=-\sigma_B \equiv \sigma$, we have
\begin{eqnarray}
    C_{n,n-1}^{(\sigma,-\sigma)}&&\hspace{-7mm}=
    \nonumber \\ &&\hspace{-13mm}
    \left\{s|s_{A1}\in C_{n-1,f}^{(\sigma,\sigma)};s_{B1}\in C_{n-1,n-2}^{(-\sigma,\sigma)};\sigma_1=\sigma;\right.
    \nonumber \\ &&\hspace{-8mm}
    \left.s_{A2}\in C_{n-1,n-2}^{(\sigma,-\sigma)};s_{B2}\in C_{n-1,f}^{(-\sigma,-\sigma)};\sigma_2=-\sigma\right\}
    \nonumber \\ &&\hspace{-16mm}
    +\left\{s|s_{A1}\in C_{n-1,n-2}^{(\sigma,-\sigma)};s_{B1}\in C_{n-1,f}^{(-\sigma,-\sigma)};\sigma_1=-\sigma;\right.
    \nonumber \\ &&\hspace{-8mm}
    \left.s_{A2}\in C_{n-1,f}^{(\sigma,\sigma)};s_{B2}\in C_{n-1,n-2}^{(-\sigma,\sigma)};\sigma_2=\sigma\right\}.
\label{eq:34}
\end{eqnarray}
The first and second terms in the right-hand sides of eqs.~(\ref{eq:33}) and (\ref{eq:34}) are graphically shown in Fig.~\ref{fig:f8}.

\noindent
We denote the number of elements in a set $X$ as $|X|$; then, eqs.~(\ref{eq:33}) and (\ref{eq:34}) result in
\begin{equation}
    |C_{n,n-1}^{(\sigma,\sigma)}|=|C_{n,n-1}^{(\sigma,-\sigma)}|=2|C_{n-1,n-2}^{(\sigma,-\sigma)}|^2,
\label{eq:35}
\end{equation}
where we have used the relations, $|C_{n-1,f}^{(\sigma,\sigma)}|=1$ and $|C_{n-1,n-2}^{(\sigma,-\sigma)}|=|C_{n-1,n-2}^{(-\sigma,\sigma)}|$.
From eq. (\ref{eq:35}), $|C_{n,n-1}^{(\sigma_A,\sigma_B)}|$ is independent of the surface spin state.
Therefore, using $|C_{n,n-1}^{(\sigma_A,\sigma_B)}|=2^{a_n}$, we rewrite eq.~(\ref{eq:35}) as
\begin{equation}
    a_n=2a_{n-1}+1,
\end{equation}
which gives
\begin{equation}
    a_n=2^{n-1}-1,
\end{equation}
where we have used $a_2=1$. 

Finally, for a sector with $l=1,\cdots, n-2$, we obtain
\begin{eqnarray}
    C_{n,l}^{(\sigma_A,\sigma_B)}&&\hspace{-7mm}=
    \nonumber \\ &&\hspace{-18mm}
    \left\{s|s_{A1}\in C_{n-1,l}^{(\sigma_A,\sigma_1)};s_{B1}\in C_{n-1,l}^{(\sigma_B,\sigma_1)};\sigma_1=\uparrow,\downarrow;\right.
    \nonumber \\ &&\hspace{-13mm}
    \left.s_{A2}\in C_{n-1,l}^{(\sigma_A,\sigma_2)};s_{B2}\in C_{n-1,l}^{(\sigma_B,\sigma_2)};\sigma_2=-\sigma_1\right\}.\;\;\;
\label{eq:cnl}
\end{eqnarray}
The total number of elements in this set is
\begin{eqnarray}
    |C_{n,l}^{(\sigma_A,\sigma_B)}|&&\hspace{-7mm}=2|C_{n-1,l}^{(\sigma_A,\uparrow)}||C_{n-1,l}^{(-\sigma_A,\uparrow)}|
    \nonumber \\ &&\hspace{-4mm}
    \times |C_{n-1,l}^{(\sigma_B,\uparrow)}||C_{n-1,l}^{(-\sigma_B,\uparrow)}|,
\label{eq:41}
\end{eqnarray}
which shows that $|C_{n,l}^{(\sigma_A,\sigma_B)}|$ does not depend on $(\sigma_A,\sigma_B)$.
Using $|C_{n,l}^{(\sigma_A,\sigma_B)}|=2^{b_{n,l}}$, eq.~(\ref{eq:41}) gives
\begin{equation}
    b_{n,l}=4b_{n-1,l}+1.
\end{equation}
We use this relation with $b_{l+1,l}=a_{l+1}=2^l-1$ to obtain
\begin{equation}
    b_{n,l}=4^{n-l-1}\left(2^l-\frac{2}{3}\right)-\frac{1}{3}.
\end{equation}

Summarizing the above results, we find that
\begin{equation}
    \log |C_{n,l}^{(\sigma_A,\sigma_B)}|=\left[4^{n-l-1}\left(2^l-\frac{2}{3}\right)-\frac{1}{3}\right]\log 2
\label{eq:46}
\end{equation}
for $l=1,2,\cdots,n-1$.

\section{Residual Entropy at Phase Boundary}

In order to calculate $S(\alpha_l)$, we write a set of internal configurations of lowest-energy states with the surface spin $(\sigma_A,\sigma_B)$ 
on the $n$-stage lattice with the diagonal interaction $\alpha_l$ as $C_{n}^{(\sigma_A,\sigma_B)}(\alpha_l)$.

Assuming $\alpha=\alpha_l$, we focus on the $(l+2)$-stage lattice.
Because $E_{l+2}(\alpha_l)=E_{l+2,l}=E_{l+2,l+1}$, we obtain $C_{l+2,l}^{(\sigma_A,\sigma_B)} \subset C_{l+2}^{(\sigma_A,\sigma_B)}(\alpha_l)$
and $C_{l+2,l+1}^{(\sigma_A,\sigma_B)} \subset C_{l+2}^{(\sigma_A,\sigma_B)}(\alpha_l)$.
Note that the set $C_{l+2,l}^{(\sigma_A,\sigma_B)}$ is constructed from four elements in $\sum_{\sigma,\sigma^{\prime}}C_{l+1,l}^{(\sigma,\sigma^{\prime})}$,
and another set $C_{l+2,l+1}^{(\sigma_A,\sigma_B)}$ is constructed from two elements in $\sum_{\sigma,\sigma^{\prime}}C_{l+1,l}^{(\sigma,\sigma^{\prime})}$
and  two elements in $\sum_{\sigma,\sigma^{\prime}}C_{l+1,f}^{(\sigma,\sigma^{\prime})}$.

\begin{figure}[h]
\begin{center}
\includegraphics[width=1\linewidth]{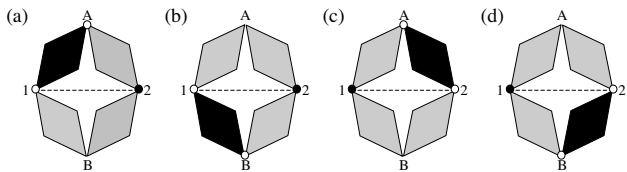}
\end{center}
\caption{Schematic representation of spin configurations in 
$\Delta C_{l+2}^{(\sigma_A,\sigma_B)}(\alpha_l)$. 
Open and closed circles indicate that the directions of those 
spins are opposite to each other.
}
\label{fig:f9}
\end{figure}

Now, we consider 
$\Delta C_{l+2}^{(\sigma_A,\sigma_B)}(\alpha_l)\equiv C_{l+2}^{(\sigma_A,\sigma_B)}(\alpha_l)-C_{l+2,l}^{(\sigma_A,\sigma_B)}-C_{l+2,l+1}^{(\sigma_A,\sigma_B)}$.
Because $E_{l+2,l}=E_{l+2,l+1}$ at $\alpha=\alpha_l$, 
we notice that there are other lowest-energy states with three elements in $\sum_{\sigma,\sigma^{\prime}}C_{l+1,l}^{(\sigma,\sigma^{\prime})}$
and one element in $\sum_{\sigma,\sigma^{\prime}}C_{l+1,f}^{(\sigma,\sigma^{\prime})}$:
\begin{eqnarray}
    \Delta C_{l+2}^{(\sigma_A,\sigma_B)}(\alpha_l)&&\hspace{-7mm}=
    \nonumber \\ &&\hspace{-25mm}
    \left\{s|s_{A1}\in C_{l+1,f}^{(\sigma_A,\sigma_A)};s_{B1}\in C_{l+1,l}^{(\sigma_B,\sigma_A)};\sigma_1=\sigma_A;\right.
    \nonumber \\ &&\hspace{-20.5mm}
    \left.s_{A2}\in C_{l+1,l}^{(\sigma_A,-\sigma_A)};s_{B2}\in C_{l+1,l}^{(\sigma_B,-\sigma_A)};\sigma_2=-\sigma_A\right\}
        \nonumber \\ &&\hspace{-28mm}
    +\left\{s|s_{A1}\in C_{l+1,l}^{(\sigma_A,\sigma_B)};s_{B1}\in C_{l+1,f}^{(\sigma_B,\sigma_B)};\sigma_1=\sigma_B;\right.
    \nonumber \\ &&\hspace{-20.5mm}
    \left.s_{A2}\in C_{l+1,l}^{(\sigma_A,-\sigma_B)};s_{B2}\in C_{l+1,l}^{(\sigma_B,-\sigma_B)};\sigma_2=-\sigma_B\right\}
        \nonumber \\ &&\hspace{-28mm}
    +\left\{s|s_{A1}\in C_{l+1,l}^{(\sigma_A,-\sigma_A)};s_{B1}\in C_{l+1,l}^{(\sigma_B,-\sigma_A)};\sigma_1=-\sigma_A;\right.
    \nonumber \\ &&\hspace{-20.5mm}
    \left.s_{A2}\in C_{l+1,f}^{(\sigma_A,\sigma_A)};s_{B2}\in C_{l+1,l}^{(\sigma_B,\sigma_A)};\sigma_2=\sigma_A\right\}
        \nonumber \\ &&\hspace{-28mm}
    +\left\{s|s_{A1}\in C_{l+1,l}^{(\sigma_A,-\sigma_B)};s_{B1}\in C_{l+1,l}^{(\sigma_B,-\sigma_B)};\sigma_1=-\sigma_B;\right.
    \nonumber \\ &&\hspace{-20.5mm}
    \left.s_{A2}\in C_{l+1,l}^{(\sigma_A,\sigma_B)};s_{B2}\in C_{l+1,f}^{(\sigma_B,\sigma_B)};\sigma_2=\sigma_B\right\}.
\label{eq:49}
\end{eqnarray}
Schematic representations of the elements in the first, second, third, and fourth sets in the right-hand side of eq.~(\ref{eq:49}) 
are respectively shown in Figs.~\ref{fig:f9}(a)--(d).

Equation~(\ref{eq:49}) gives
\begin{equation}
    |\Delta C_{l+2}^{(\sigma_A,\sigma_B)}(\alpha_l)|=2^{3\cdot 2^l-1},
\end{equation}
and thus, we obtain
\begin{equation}
    |C_{l+2}^{(\sigma_A,\sigma_B)}(\alpha_l)|=2^{4\cdot 2^l-3}+2^{2^{l+1}-1}+2^{3\cdot 2^l-1}.
\end{equation}

For the $n$-stage lattice with $n>l+2$, using
\begin{equation}
    |C_{n}^{(\sigma_A,\sigma_B)}(\alpha_l)|=2^{d_n},
\end{equation}
we obtain
\begin{equation}
    d_n=4d_{n-1}+1.
\end{equation}
The recursion relation with the initial condition
\begin{equation}
    d_{l+2}=4\cdot 2^l-3+\frac{\log(1+2^{2-2^{l+1}}+2^{2-2^l})}{\log 2}
\end{equation}
gives
\begin{equation}
    d_n=4^{n-l-2}
    \left[4\cdot 2^l-\frac{8}{3}+\frac{\log(1+2^{2-2^{l+1}}+2^{2-2^l})}{\log 2}\right]
    -\frac{1}{3}.
\end{equation}
Thus, the entropy at $\alpha=\alpha_l$ is given by
\begin{eqnarray}
    S(\alpha_l)&&\hspace{-7mm}
    =\lim_{n\rightarrow\infty}\frac{1}{N_n}\log 4\cdot 2^{d_n}
    \nonumber\\
    &&\hspace{-7mm}=S_{\infty,l}+\frac{3}{8\cdot 4^l}
    \log\left[1+4(2^{-2^l}+4^{-2^l})\right].
    \hspace{3mm}
\end{eqnarray}


\end{document}